# Ambipolar Surface Conduction in Ternary Topological Insulator $Bi_2(Te_{1-x}Se_x)_3$ Nanoribbons

ZhenHua Wang,[†,‡] Richard L. J. Qiu,[‡] Chee Huei Lee,[‡] ZhiDong Zhang[†] and Xuan P. A. Gao[*,‡]

[†]Shenyang National Laboratory for Materials Science, Institute of Metal Research, and International Centre for Materials Physics, Chinese Academy of Sciences, 72 Wenhua Road, Shenyang 110016, People's Republic of China

[‡]Department of Physics, Case Western Reserve University, Cleveland, Ohio 44106, United States

[*]To whom correspondence should be addressed. Email: xuan.gao@case.edu

**ABSTRACT** We report the composition and gate voltage induced tuning of transport properties in chemically synthesized $Bi_2(Te_{1-x}Se_x)_3$ nanoribbons. It is found that increasing Se concentration effectively suppresses the bulk carrier transport and induces semiconducting behavior in the temperature dependent resistance of $Bi_2(Te_{1-x}Se_x)_3$ nanoribbons when $x$ is greater than ~10%. In $Bi_2(Te_{1-x}Se_x)_3$ nanoribbons with $x$ ~20%, gate voltage enables ambipolar modulation of resistance (or conductance) in samples with thickness around or larger than 100nm, indicating significantly enhanced contribution in transport from the gapless surface states.

Three-dimensional (3D) topological insulators (TIs) with both metallic surface states and insulating bulk states have attracted enormous research attention. The metallic surface states in 3D TIs consist of a single Dirac cone at the $\Gamma$ point which gives rise to interesting surface transport phenomena.[1-3] Binary topological insulators $Bi_2Te_3$, $Bi_2Se_3$ and $Sb_2Te_3$ have been confirmed to have robust surface states by surface sensitive probes such as angle resolved photoemission spectroscopy (ARPES)[4-7] and scanning tunneling microscopy/spectroscopy (STM/STS).[8,9] The charge carriers originating from surface states also have been identified by magneto-transport experiments.[10-14] In addition, electrical gating was used as a means to manipulate the gapless surface state conduction in 3D TIs in $Bi_2Te_3$ and $Bi_2Se_3$.[13,15-18] However, the topological surface contribution in transport is hindered by residual bulk carriers from environmental doping or crystal defects.[10,19,20] Therefore, to exploit the surface transport properties of topological insulators, it is crucial to achieve a bulk-insulating state in a topological insulator material.

Recently, the ternary and quaternary topological insulators have been synthesized to maximize the surface properties. The Shubnikov-de Haas (SdH) oscillations coming from the surface states were observed in bulk $Bi_{2-x}Sb_xSe_3$,[10] Cd-doped $Bi_2Se_3$[21] and $Bi_2Te_2Se$ crystals[22] and nanopletes.[23] However, no SdH oscillation was observed in $Bi_{2-x}Ca_xSe_3$[24] due to the disorder from the low level substitution of $Ca^{2+}$ for $Bi^{3+}$. $Bi_2Te_2Se$ presented a high resistivity and a variable-range hopping behavior, which is crucial to achieve the surface transport properties.[22] Cd-doped $Bi_2Se_3$ showed a transition from n-type to p-type conversion upon increasing Se vacancies by post annealing.[21] The solid solution $Bi_{2-x}Sb_xTe_{3-y}Se_y$ showed a series of "intrinsic" compositions and presented a maximally bulk-insulating behavior, which can achieve a surface-dominated transport.[25,26] The SdH oscillation of both Dirac holes and electrons was observed in the compound $Bi_{1.5}Sb_{0.5}Te_{1.7}Se_{1.3}$.[26] Besides studies on bulk materials, topological insulator nanoribbons and nanoplates are expected to have significantly enhanced surface conduction because of their large surface-to-volume ratios.[11] $(Bi_xSb_{1-x})_2Te_3$ nanoplates with thickness less than 10nm have been shown to exhibit ambipolar field effect.[27] Alternatively, the Fermi level of $Bi_2Te_3$ nanoplates could be tuned by Na doping and enhanced surface states conduction was observed, in which the ambipolar surface transport could be achieved by applying a back-gate voltage.[28] $Bi_2(Se_xTe_{1-x})_3$ is a topological insulator for all atomic ratio $x$, similar to $(Bi_xSb_{1-x})_2Te_3$.[27,29,30] In $Bi_2(Se_xTe_{1-x})_3$ nanoribbons and nanoplates with high carrier density, two-dimensional (2D) weak anti-localization was clearly observed.[30] Similar 2D weak anti-localization effect was also observed in $Bi_2Te_2Se$ nanoplates.[31] However, ambipolar field effect has never been reported in the ternary $Bi_2(Se_xTe_{1-x})_3$ TI materials. Here, we report the synthesis and transport studies of $Bi_2(Te_{1-x}Se_x)_3$ nanoribbons. For nanoribbons with thickness ~100-300nm, we show that increasing the Se concentration, $x$, gradually tunes the material from metallic to semiconducting. Moreover, ambipolar field effect from surface state conduction was achieved by applying a back gate voltage for semiconducting compound with $x$~20%. The thickness of nanoribbons is also found to be important for observing enhanced ambipolar field effect.

## Results and Discussion

Vapor-liquid-solid (VLS) process was used to synthesize $Bi_2(Te_{1-x}Se_x)_3$ nanoribbons in a vapor transport setup (see Methods section). Figure 1A shows scanning electron microscope (SEM) images of the as synthesized ribbons, the length of the ribbons ranges from several μm to tens of μm and the width ranges from a few hundred nm to a few μm. The wide ribbons display 120° facet angles reflecting the crystal structure. The concentrations of Bi, Te and Se in $Bi_2(Te_{1-x}Se_x)_3$ nanoribbons were estimated by energy-dispersive X-ray spectroscopy (EDS) on individual nanoribbons. An example of EDS data corresponding to $Bi_2(Te_{1-x}Se_x)_3$ nanoribbon with x=20±5% is shown in Figure 1B. To investigate the transport properties and explore the topological surface conduction in $Bi_2(Te_{1-x}Se_x)_3$ nanoribbons, back-gated field effect transistor (FET) devices were fabricated on doped Si substrates with 300nm thick $SiO_2$ on surface and measured in a cryostat (see Methods Section).[14]

Temperature dependent resistance measurements of $Bi_2(Te_{1-x}Se_x)_3$ nanoribbons with different $x$ show that substituting Te with Se can indeed suppress the residual bulk carriers and tune the Fermi level into bulk bandgap, in a similar way to the studies in bulk crystals.[25] In Figure 2, we compare $R$ vs. $T$ for a series of $Bi_2(Te_{1-x}Se_x)_3$ nanoribbons with different $x$. To compare the temperature dependence directly, the resistances of different samples are normalized by the values at 300K. A clear trend can be seen in Figure 2 that as more Se is incorporated into the $Bi_2Te_3$, the material turns from metallic into semiconducting at $x > \sim 10\%$. Atomic force microscope (AFM) measurements revealed that the thickness of device #4, #5 and #6 is 105±2, 320±2 and 205±2 nm, respectively. When Se concentration is 20±5%, $R(T)$ shows an insulating behavior, but it saturates below ~75 K due to the metallic surface transport as well as the bulk impurity-band transport.[25,26] The nonmetallic $R$ vs. $T$ observed in these nanoribbons is consistent with those reported in the bulk crystals of $Bi_{2-x}Sb_xTe_{3-y}Se_y$,[25,26] and the similar nonmetallic properties were also observed in Na-doped $Bi_2Te_3$ nanoplates and $Bi_2(Se_xTe_{1-x})_3$ nanoribbons/plates.[28,30]

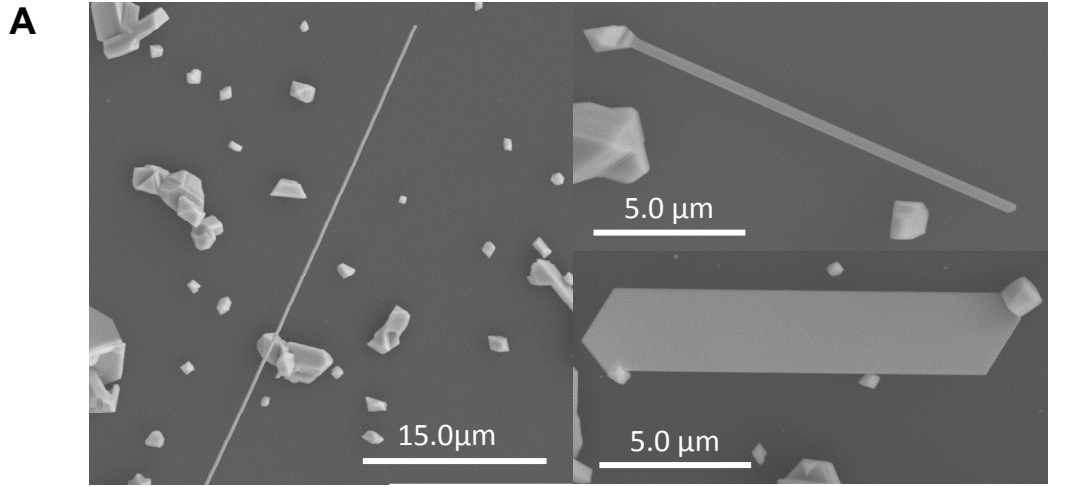

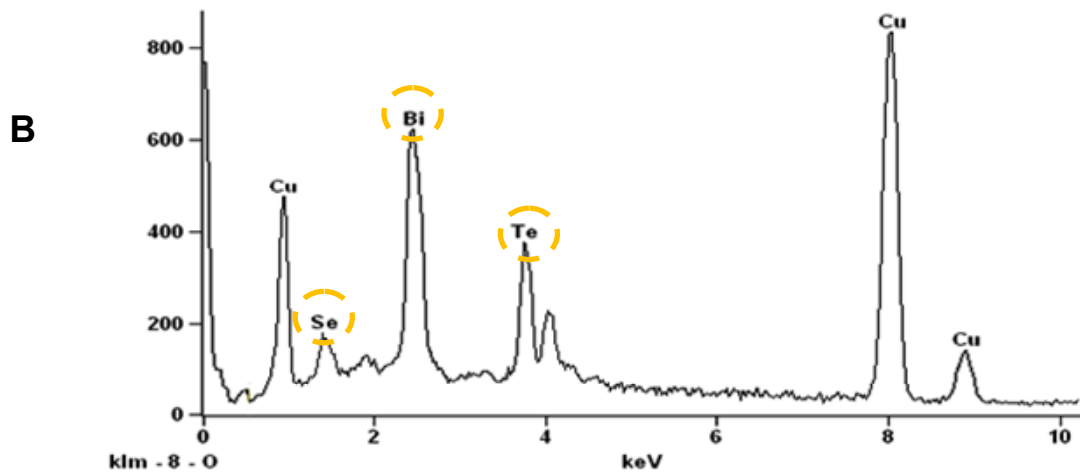

| Element Line | Weight % | Weight % Error | Atom % | Atom % Error |
|---|---|---|---|---|
| Cu K | 57.32 | +/- 0.74 | 76.46 | +/- 0.99 |
| Se K | 3.07 | +/- 0.74 | 3.30 | +/- 0.79 |
| Te L | 16.16 | +/- 0.75 | 10.73 | +/- 0.50 |
| Bi M | 23.45 | +/- 0.65 | 9.51 | +/- 0.26 |
| Total | 100.00 | | 100.00 | |

Figure 1. Morphological and composition data of $Bi_2(Te_{1-x}Se_x)_3$ nanoribbons. (A) SEM images of $Bi_2(Te_{1-x}Se_x)_3$ nanoribbons with length from several μm to tens of μm and width from several hundred nm to several μm. (B) EDX data of a $Bi_2(Te_{1-x}Se_x)_3$ nanoribbon with x≈20±5%. The Cu peaks originated from the Cu thin film substrates.

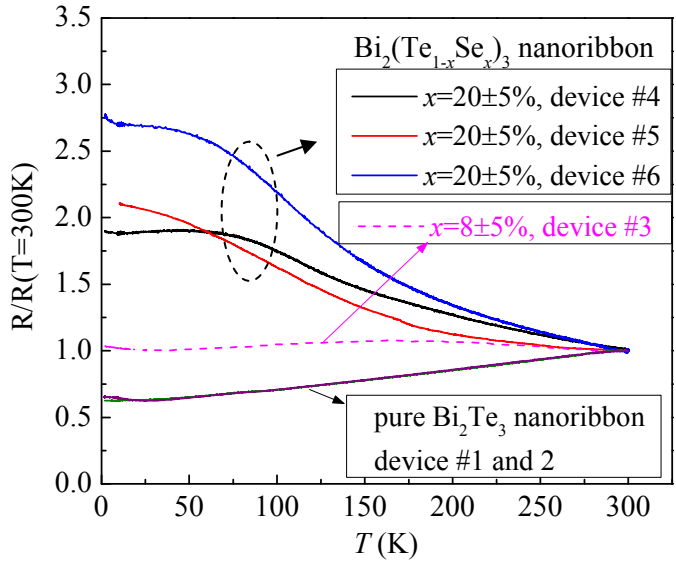

Figure 2. Composition tuned temperature dependent transport in $Bi_2(Te_{1-x}Se_x)_3$ nanoribbons. The dependence of resistance with temperature showing the transport property changes from metallic into semiconducting as the concentration of Se increases in $Bi_2(Te_{1-x}Se_x)_3$ nanoribbons. The resistances at 300K are 77Ω, 970Ω, 4440Ω, 6120Ω, 49.2kΩ, and 4.3kΩ for device #1-6.

Since the topological surface states are gapless, ambipolar conduction in the gate voltage tuned devices is taken as a signature of TI surface transport.[16, 17, 27] For $Bi_2(Te_{1-x}Se_x)_3$ nanoribbons or $Bi_2Te_2Se$ nanoplates, either very high electron density was observed[30] or only the n-type conduction was achieved in the gating experiment, meaning the Fermi level was still above the Dirac point.[31] In our experiments, we found that with the bulk carrier conduction suppressed by the incorporation of Se, nanoribbons of ternary $Bi_2(Te_{1-x}Se_x)_3$ could start to exhibit ambipolar field effect in samples with thickness ~100nm or even larger. Figure 3A-B shows the zero-field resistance as a function of the back gate voltage ($V_g$) at various temperatures for device #4, where $x$ is ~20% and the sample thickness is ~105nm. It is interesting to note that the gate modulated resistance showed distinct behaviors at different temperature. At 300 K, the resistance increased with decreasing the gate voltage, exhibiting a clear n-type feature. When the temperature was decreased to 200 K, the device kept weak n-type characteristic with high gate voltage, however the resistance started to decrease when the applied gate voltage was below about +20 V. This behavior indicates a transition from n-type to p-type conduction. When the temperature dropped further down to 2 K, an ambipolar characteristic became more evident in the $R(V_g)$ curve and the turning point moved to ~-20V, as shown in Figure 3B. This behavior in the temperature effect on the ambipolar field effect can be understood by considering the relative contributions from surface and bulk states at different $T$: since the semiconducting $R(T)$ of $Bi_2(Te_{.8}Se_{.2})_3$ samples in Figure 2 suggests that the Fermi level resides in the bulk band gap in these samples, there are exponentially more thermally excited electrons in the bulk at high temperatures which may smear out the ambipolar field effect from the gapless surface states. Further studies on the gate tuned Hall resistivity at different temperatures would be required to quantitatively distinguish the surface carrier density and mobility from the bulk states by a multi-band transport analysis, as previously done on bulk or exfoliated flakes of topological insulator materials. [17, 21, 22, 32] Unfortunately, making Hall measurement has been difficult for our nanowires/nanoribbons. Such measurement and analysis will be a subject of future study.

In addition to ambipolar field effect, we observed that 2D weak anti-localization effect also evolves with the gate voltage. Figure 3c shows the conductivity $\sigma_{xx}$ in units of $e^2/h$ *versus* perpendicular magnetic field ($B$) at 2 K at various $V_g$. The sharply peaked behavior near $B=0$ indicates weak antilocalization effect. We fit the conductivity data to 2D weak antilocalization model $\Delta\sigma_{xx} = \alpha e^2/(2\pi^2\hbar)[\ln(B_\varphi/B) - \psi(1/2 + (B_\varphi/B))]$, where $B_\varphi = \hbar/(4eL_\varphi^2)$ is a characteristic field defined by the dephasing length $L_\varphi$, $\psi$ is the digamma function and $\alpha$ is a constant.[17, 18, 33] Correlating with the crossover from n- to p-type conduction at $V_g$~-20V, we observe that $\alpha$ increases as $V_g$ becomes smaller than -20V. Previous measurements on molecular beam epitaxy grown $Bi_2Se_3$ thin films with thickness of 20nm or thinner showed that $\alpha$ increased from ~0.5 or 0.7 toward 1, concomitant with the ambipolar field effect in the gate tuned conductance.[18, 33] It is known that both conventional 2D films with strong spin-orbit coupling and a single layer of TI surface can contribute weak antilocalization correction with $\alpha$ =1/2. So this gate tuned change of $\alpha$ from 0.5 to 1 was attributed to the gate induced separation of bulk states from surface states when one of the TI surfaces was gated to the Dirac point, leaving the bulk states and the top TI surface each contributing 0.5 to the total $\alpha$.[18, 33] Although we have obtained a similar increasing trend of $\alpha$ at $V_g$<-20V accompanying the ambipolar conduction, the magnitude of enhancement in $\alpha$ is smaller, and $\alpha$ itself (~0.3-0.34) is smaller than 0.5. This discrepancy in the quantitative magnitude of $\alpha$ may lie in the fact that for our sample, the conductivity $\sigma_{xx}$ is only 3-4×$e^2$/h. Strictly speaking, the weak antilocalization fitting only works in the diffusive metallic transport regime where $\sigma_{xx}$ >>$e^2$/h. For $\sigma_{xx}$ comparable to quantum conductance $e^2$/h, it is reasonable to expect that the standard weak antilocalization model does not give a quantitatively accurate fitting result although it captures the qualitative physics.

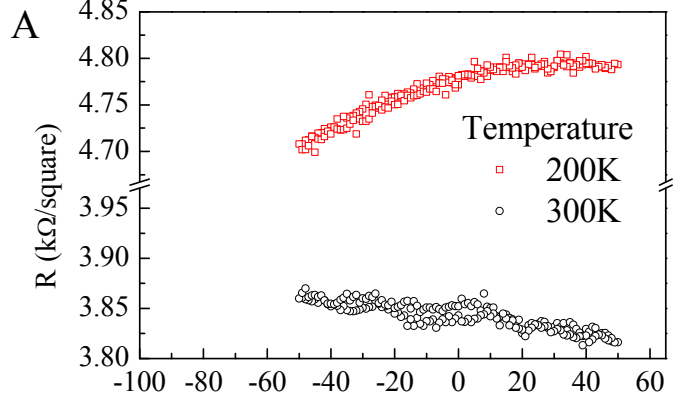
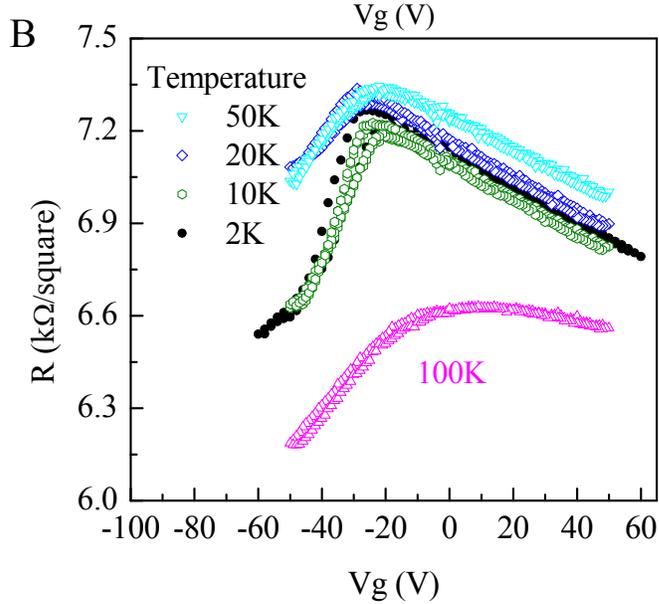
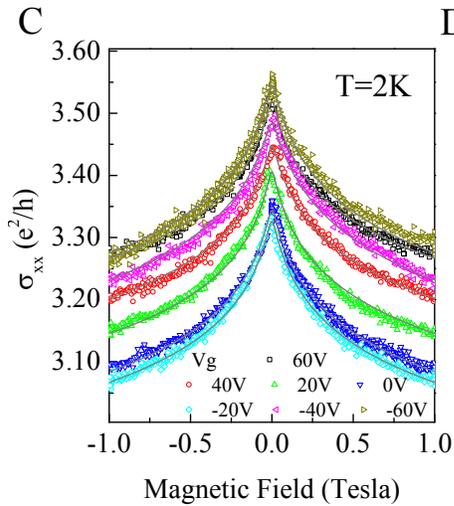
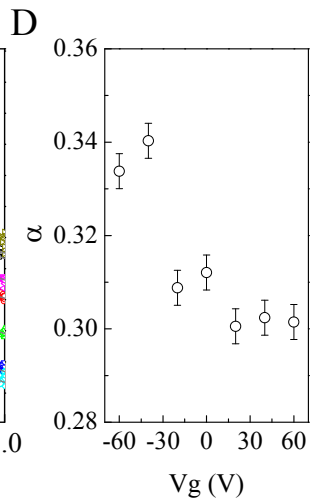

Figure 3. Gate voltage tuned ambipolar surface conduction and weak anti-localization in $Bi_2(Te_{0.8}Se_{0.2})_3$ nanoribbon. (A, B) The dependence of resistance per square *vs*. gate voltage at different temperatures for a 105nm thick $Bi_2(Te_{0.8}Se_{0.2})_3$ nanoribbon (device #4) which shows ambipolar behavior at *T*<200K. (C) Magneto-conductivity of the device at different gate voltage showing weak antilocalization effect. The symbols are data and solid lines are fits. (C) The fitting parameter α in the weak antilocalization fit *vs*. gate voltage.

It is remarkable that clear ambipolar field effect is observed in nanoribbons with thickness ~100nm, attesting to the effectiveness of including Se to compensate the defect states in $Bi_2Te_3$ to achieve an insulating bulk behavior. Previously, gate voltage induced ambipolar conduction was observed in $Bi_2Se_3$ flakes or films with thickness less than 50nm.[16-18,24] In $(Bi_xSb_{1-x})_2Te_3$ nanoplates, the ambipolar field effect was very weak once the plate thickness increased above 9nm.[27] Indeed, comparing our $Bi_2(Se_xTe_{1-x})_3$ nanoribbons with the same *x* but different thickness we found thicker samples exhibited weaker ambipolar field effect, due to the more dominant bulk contribution to the overall conductance. For 105nm thick device #4, the ambipolar effect can be clearly seen even at 100 K, as shown in Figure 3B. But for thicker nanoribbon devices #5 (thickness ~205nm) and #6 (thickness ~320nm), the dependence of *R* on $V_g$ shows a weaker tunability, as shown in Figure 4. Moreover, the ambipolar surface conduction effect disappeared at 20 and 10 K for device #5 and #6 respectively. In addition, it is noted that the gate voltage for the transition from n- to p-type conduction is about -20V for 105nm thick device #4. However, with the thickness increased, this gate voltage is increased to about -60 V and -70 V for device #5 and device #6. Therefore, it is important to fabricate devices with small thickness to enhance the ratio between surface conductance and bulk conductance for better characterization of topological surface transport.

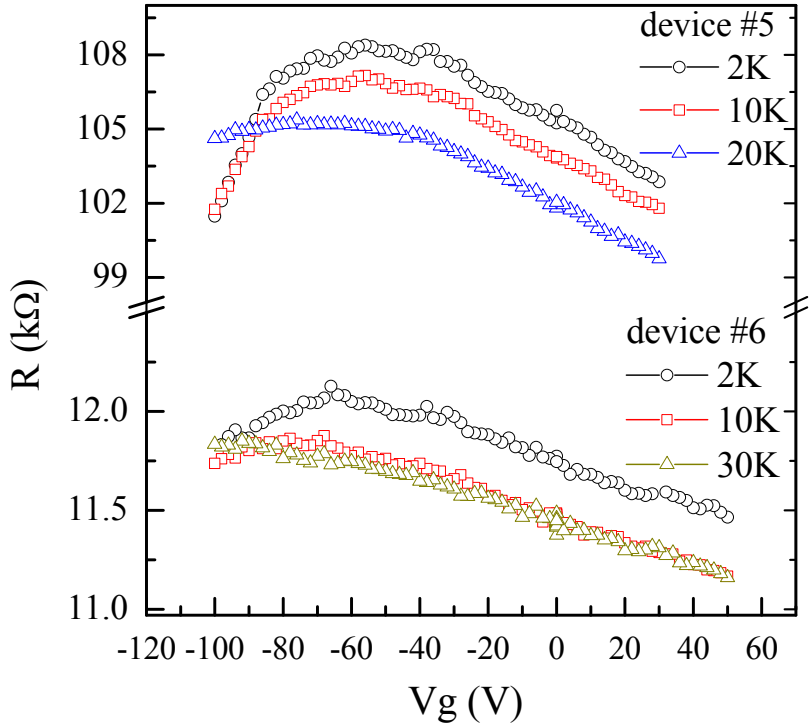

Figure 4. Ambipolar conduction in $Bi_2(Te_{1-x}Se_x)_3$ nanoribbons with large thickness. Gate voltage tuned ambipolar conduction at various temperatures in $Bi_2(Te_{1-x}Se_x)_3$ nanoribbons with $x\sim0.2$ and the thickness about 205 nm (device #5) and 320 nm (device #6).

## CONCLUSION

In summary, we realized efficient suppression of bulk carrier transport in $Bi_2(Se_xTe_{1-x})_3$ nanoribbons/nanowires with increasing Se concentration $x$. Semiconducting temperature dependent resistivity was observed in $Bi_2(Se_xTe_{1-x})_3$ nanoribbons/nanowires with thickness of 100-300nm at $x\sim20\%$, indicating that the Fermi level was tuned into bulk band gap. Moreover, as a result of reduced bulk conduction, ambipolar field effect from the gapless surface states was observed in $Bi_2(Se_{.8}Te_{.2})_3$ nanoribbons/nanowires with thickness more than 100nm. This study illustrates the usefulness of composition tuned ternary topological insulator nanomaterials in transport studies of topological insulators.

## METHODS

Vapor-liquid-solid (VLS) growth method was used to synthesize ternary $Bi_2(Te_{1-x}Se_x)_3$ nanoribbons, using 20 nm gold nanoparticles (Ted Pella, Inc) as metal catalyst on Si substrates. Pure $Bi_2Te_3$ and $Bi_2Se_3$ powders (99.999%, Alfa Aesar) were used as precursor. $Bi_2Te_3$ was placed in the center of a one inch diameter quartz tube inside a furnace (Lindberg/Blue M) and $Bi_2Se_3$ was placed at different sites far from the center at low temperature area, the atomic ratio of Te/Se in these nanoribbons was controlled by changing the distance of $Bi_2Te_3$ and $Bi_2Se_3$ precursors. The growth conditions for $Bi_2(Te_{1-x}Se_x)_3$ nanoribbons were similar to the conditions for $Bi_2Se_3$ nanoribbons.[14] The concentrations of Bi, Te and Se in $Bi_2(Te_{1-x}Se_x)_3$ nanoribbons were estimated by energy-dispersive X-ray spectroscopy (EDS) equipped in HITACHI S4500 SEM. Back-gated $Bi_2(Te_{1-x}Se_x)_3$ nanoribbon field effect transistor (FET) devices were fabricated by standard photolithography patterning and liftoff process, with doped Si wafers as substrates. The Si substrates had 300nm thick $SiO_2$ on surface as gate dielectric. The transport properties of $Bi_2(Te_{1-x}Se_x)_3$ nanoribbon devices were measured in a Quantum Design PPMS (physical property measurement system) using low frequency lock-in technique. The resistance $R$ was measured either in a four-probe [14] or two-probe configuration. We found that the contact resistance was small, and four-probe measurement gave result similar to that of a simple two-probe configuration.

**Acknowledgments** Z.H.W. acknowledges China Scholarship Council for a scholarship supporting her visit to CWRU. X. P. A. G acknowledges the NSF CAREER Award program (grant number DMR-1151534) for financial support of research at CWRU. Z.D.Z acknowledges the National Natural Science Foundation of China with Grant No. 50831006.

TOC

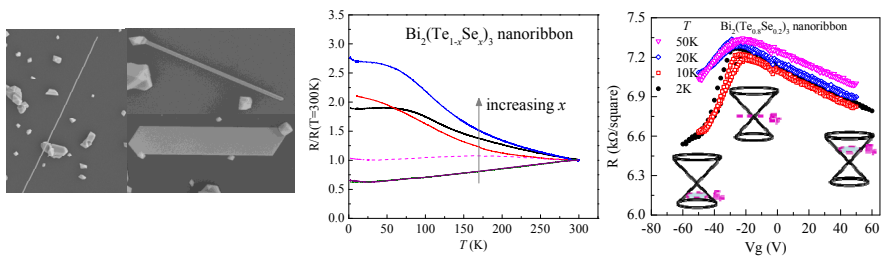